\documentclass[aps,prl,showpacs,showkeys,twocolumn]{revtex4-1}
\usepackage{graphicx,hyperref,amsmath,amsfonts}
\usepackage{epstopdf,color,bm,multirow,rotating,soul}

\begin{document}
\title{Spectrally improved controllable frequency comb quantum memory}
\author{N.S. Perminov$^{1,2}$, D.Yu. Tarankova$^{3}$ and S.A. Moiseev$^{1,2, *}$}
\affiliation{$^{1}$Kazan Quantum Center, Kazan National Research Technical University n.a. A.N.Tupolev-KAI, 10 K. Marx, Kazan 420111, Russia}
\affiliation{$^{2}$Zavoisky Physical-Technical Institute, SSS FRC KSC RAS, 10/7 Sibirsky Tract, Kazan 420029, Russia}
\affiliation{$^{3}$Institute of Radio-Electronics and Telecommunications, Kazan National Research Technical University n.a. A.N.Tupolev-KAI, 10 K. Marx, Kazan 420111, Russia}
\email{s.a.moiseev@kazanqc.org}
\pacs{ 03.67.-a, 03.67.Hk, 03.67.Ac, 84.40.Az.}
\keywords{quantum information, broadband quantum memory, controlled frequency comb, spectrum optimization, ring resonator.}

\begin{abstract}						
We propose a scheme of a universal block of broadband quantum memory consisting of three ring microresonators forming a controllable frequency comb and interacting with each other and with a common waveguide.
We find the optimal parameters of the microresonators showing the possibility of highly efficient storage of light fields on this memory block and we demonstrate the procedure for gluing several memory blocks for increasing spectral range of the composite quantum memory while maintaining high efficiency.
\end{abstract}

\date{\today}

\maketitle

\textit{Introduction.}
The dynamics of multiparticle systems possessing a large number of controllable parameters \cite{Hartmann2008,Roy2017,Noh2017} provides ample opportunities for constructing composite resonance circuits with predefined spectral properties. 
Significant progress along this way was achieved when controlling a light propagation in a system of interconnected high-Q resonators that allow creating the 
light fields distributed in space by a system of coupled resonators, called "photonic molecules" \cite{Li2017}. 
This type of multiresonator systems are successfully developed as controllable optical delay lines \cite{Heebner2002,Xia2006,Yariv1999,Melloni2010,Du2016} easily coupled into fiber-optic lines.

The considerable improvement of high-Q resonators \cite{Gorodetsky1999,Vahala2003,Kobe2017,Toth2017,Megrant2012} and their integration into multiresonator structures \cite{Armani2003} makes them interesting for use in optical and microwave quantum memory (QM) circuits \cite{EMoiseev2017,Moiseev_2017_PRA,Moiseev2018} in which resonators can have a specified periodic frequencies. 
Such schemes demonstrate the possibility of a significant increase in the operating spectral range of the QM and high quality of the resonators makes it possible to considerable enhance the constant coupling both between neighboring high-Q resonators and light-atoms interaction in the resonators. 
The latter properties make this type of QM promising for use in circuits of a universal quantum computer \cite{Kurizki2015,Moiseev2016,Moiseev2013,Perminov2017superefficient,Perminov2017spectral,Kockum2018}.

\begin{figure}[h]
\includegraphics[width=0.45\textwidth]{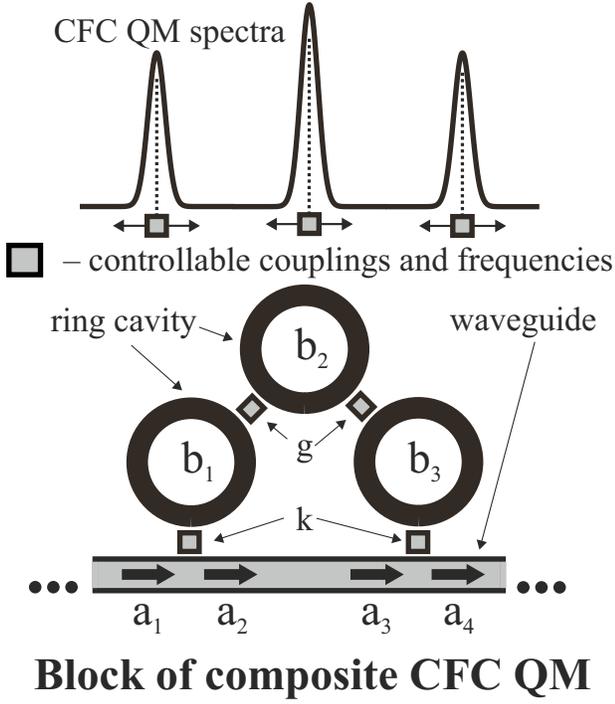}
\caption{Principal scheme of the block of composite CFC QM from ring microresonators.}
\label{fig_1}
\end{figure}

In this paper, we suggest a new composite QM scheme consisting of series connected compact QM blocks (see Fig. \ref{fig_1}), each of them contains several resonators coupled to each other and with a common waveguide. 
We find the optimal parameters of the resonators and their connections in the block and we develop the procedure for integrating the blocks into the common QM scheme providing high quantum efficiency of light field storage in the spectral range by an order of magnitude greater than the line width of the individual resonators. 
The possible implementation of the proposed scheme at room temperature is also discussed.

\textit{Physical model.}
The initial idea of the scheme under consideration is based on the photon echo QM approach \cite{Moiseev2001,Moiseev2007,Tittel2009} in a variant using resonant atoms with a periodic spectral structure of the inhomogeneous broadening of the absorption line, which is known as the AFC protocol \cite{Riedmatten2008}, and also on the realization of this approach in the optimal resonator \cite{Moiseev2010,Afzelius2010}. 
In contrast to \cite{Riedmatten2008,Sabooni2013_1,Jobez2014,Akhmedzhanov2016}, in our QM scheme, instead of the atomic system, we use a system of microresonators optimally connected  with a common waveguide \cite{EMoiseev2017,Moiseev_2017_PRA} or a waveguide resonator \cite{Moiseev2018}, which opens the possibility of creating a broadband high-performance interface on a finite number of resonators.

Each block of the QM circuit with controllable frequency comb (CFC), shown in Fig. \ref{fig_1}, consists of three connected circular microresonators having different frequencies and interacting with the common waveguide through the first and third microresonators. 
Inside the block, the geometric arrangement of the microresonators along the waveguide and with respect to each other is chosen so that when the input light passes from the first resonator to the third through the waveguide or through the second microresonator, the optical path is a multiple of the whole number of half-wave lengths and there is no additional phase shift. 
We also assume that irreversible losses in resonators are negligible at the times under consideration and the symmetry of the spectroscopic parameters (frequencies, coupling constants of the microresonators with each other and a fiber) is satisfied with respect to the second microresonator situated inside each block.
The symmetry condition is a natural requirement to the universal QM blocks from which it would be possible to make a common composite QM with an arbitrary working spectral width. 
We study the possibilities of spectral improvement of the considered QM blocks due to optimization of their spectroscopic parameters and coupling constant, and we also consider the procedure of optimal gluing of the QM blocks, which allows increasing the width the QM bandwidth while maintaining the its high overall quantum efficiency.

Using the well-known formalism of quantum optics \cite{Walls} to describe the interaction of light fields in waveguide resonators, we obtain the following equations for the input $a_1(t),a_3(t)$ and output $a_2(t),a_4(t)$ light fields in the waveguide (Fig. \ref{fig_1}) coupled with the modes of the ring microresonators $b_1(t),b_2(t),b_3(t)$:
\begin{align}\label{eq_1}
& \nonumber \left[\partial_{t}+i\Delta_1+k_1/2\right]b_1(t)+ig_1 b_2(t)=\sqrt{k_1}a_1(t),\\
& \nonumber \left[\partial_{t}+i\Delta_2\right]b_2(t)+ig_1^* b_1(t)+ig_2^*b_3(t)=0,\\
& \nonumber \left[\partial_{t}+i\Delta_3+k_2/2\right]b_3(t)+ig_2 b_2(t)=\sqrt{k_2}a_3(t),\\
& \nonumber a_1(t)-a_2(t)=\sqrt{k_1}b_1(t),\\
& \nonumber a_3(t)-a_4(t)=\sqrt{k_2}b_3(t),\\
& a_2(t)=a_3(t),
\end{align}
where $\Delta_1=-\Delta,\Delta_2=0,\Delta_3=\Delta$ are the detunings of the microresonators from the central frequency $\nu_0$, $g_1=g,g_2=g$ are the coupling constants between the microresonators, $k_1=k,k_2=k$ are the coupling constants between the microresonators and the waveguide.

Using the Fourier transform for the field mode in Eqs.(\ref{eq_1}), we find the output field $a_4(t)$ in terms of the transfer function (TF) $S(\nu)=\tilde{a}_4(\nu)/\tilde{a}_1(\nu)$:
\begin{align}\label{eq_2}
& \nonumber S(\nu)= \exp⁡\{i \nu T(\nu)\},\\
& T(\nu)=\operatorname{Arctan}(4k\nu/[8g^2+k^2+4-4\nu^2])/\nu,
\end{align}
where $a_{1,4}(t)= [2\pi]^{-1/2} \int d\nu e^{-i\nu t} \tilde{a}_{1,4}(\nu)$, a function $T(\nu)$ is a spectral time delay (the storage duration) of the detuning $\nu$ (without loss of generality we set $\Delta=1$, i.e. here and below all quantities are given in units of $\Delta$).

\textit{Optimizing the QM block.}
As can be seen from (\ref{eq_2}), the spectral efficiency $\eta(\nu)=|S(\nu)|^2$ of the QM under study in any frequency band $\nu$ is equal to 1 and the main physical content of the memory properties is contained in the spectral properties of function $T(\nu)$ determining the accuracy of recovering the arbitrary input signal. 
It should be remembered that optimizing the recovery fidelity for a particular input signal $a_1(t)$ and optimizing the function $T(\nu)$ in a wide frequency range (for an input signal $a_1(t)$ of arbitrary shape) are the different tasks. 
The ideal infinite-broadband QM is characterized by the condition $T_{rel}(\nu)=T(\nu)/T(0)=1$ for any frequency detuning $\nu$, which can not be achieved due to the finiteness of the number of absorbers.
For a system with a finite number of absorbers, the last condition must be replaced by an approximate equality $T(\nu)\cong T(0)$, which is equivalent to optimizing the quantity $|T(\nu)-T(0)|^2$ in a wide frequency range \cite{Perminov2017superefficient,Perminov2017spectral}. 
Physically, this means that the function $T(\nu)$ has a spectral plateau in some neighborhood $\nu=0$ (the region where the function has a sufficiently small changes), which can be made more flat and wide due to the optimization and that increases the effective QM bandwidth, respectively.

\begin{figure}[h]
\includegraphics[width=0.45\textwidth]{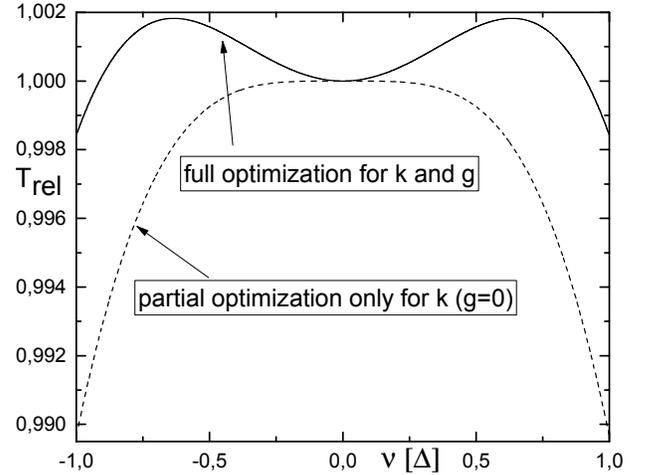}
\caption{The normalized delay time $T_{rel}$, depending on the frequency $\nu$ for the optimized QM block in the case of the switched off coupling $\{g=0,k=3,17\}$ (partial optimization is the dashed line), and the switched on coupling $\{g=0,29,k=3,47\}$ (full optimization is the solid line). The spread of the delay time for the solid curve is $0,004$ vs. $0,006$ for the dotted curve corresponding only to partial optimization.}
\label{fig_2}
\end{figure}

According to \cite{Perminov2017superefficient}, for QM with $N_0$ resonators coupling the waveguide, we use the objective function $H=\sum_{m=-N_0/2}^{N_0/2}|T(\nu=2m/N_0)|^2$ for optimization (minimization) in spectral range of the order $N_0 \Delta$ ($\Delta=1$). 
The case of 2 free varied parameters $k,g$ is called complete optimization, and the case of one free varied parameter $k$ is a partial optimization (where $g=0$ in each block so an auxiliary resonator does not interact directly with the common waveguide and it is actually absent in the system).

The results of optimization of one block of QM are shown in Fig. \ref{fig_2}, where the normalized delay time $T_{rel}$ is shown as a function of the frequency $\nu$ for a partially optimized QM block with  switched off coupling $\{g=0,k=3,17\}$ and fully optimized block with switched on coupling $\{g=0,29,k=3,47\}$. 
Time delay variation in the frequency band $[-\Delta;\Delta]$ is $0,004$ for the total optimization versus $0,006$ for the partial optimization. 
Physically, this means that the connection of an additional microresonator to the QM block \cite{EMoiseev2017} of two resonators makes it possible to reduce the spectral errors in the working bandwidth by the factor of $1,5$ or to use a $1,5$ times wider frequency band, while maintaining the previous spectral accuracy.

\textit{Procedure of gluing blocks.}
To significantly increase the spectral bandwidth of the QM, we propose optimal gluing of several QM blocks along the waveguide line where each black covers its own frequency band while the adjacent bands of QM blocks are partially overlapped in an optimal way. 
Due to the spectral gluing, the working bandwidth  can be increased by several orders of magnitude for the composite QM in comparison with the spectral width of each QM block and resonator linewidth, respectively. 
We demonstrate the optimal spectral gluing of 2 similar fully optimized QM blocks whose frequency centers are shifted on $-\delta$ and $\delta$, and we compare the obtained circuit with the previously proposed scheme \cite{EMoiseev2017} where 4 resonators are connected to a fiber whose frequencies are equidistantly detuned equal to $\Delta$ and the coupling constant of each with the waveguide is $k$.

\begin{figure}[h]
\includegraphics[width=0.45\textwidth]{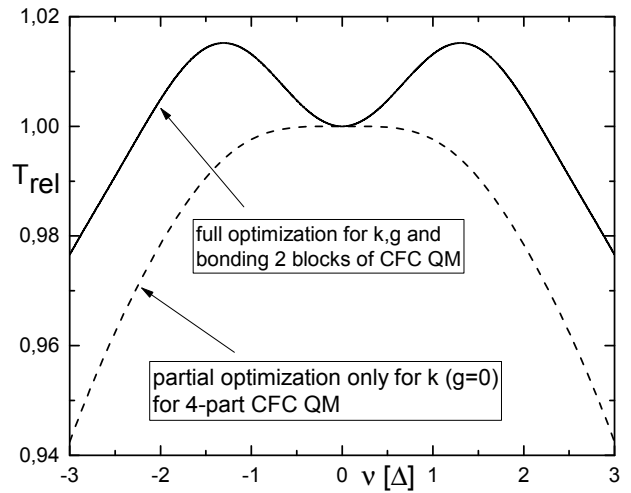}
\caption{The normalized delay time $T_{rel}$, depending on the frequency $\nu$ for optimal gluing of 2 QM blocks in the case of the switched off coupling $\{g=0,k=4,26\}$, and only 4 resonators are used (partial optimization -- dashed line), and the switched on coupling $\{\delta=2,18\}$ (full optimization -- solid line); the spread of the delay time for the dotted curve is $0,06$, and for the solid curve is $0,04$, which corresponds to the best spectral quality of the QM scheme.}
\label{fig_3}
\end{figure}

For the blocks gluing, the maximum deviation of $|T_{rel}(\nu)-1|$occurs at the midpoint of the spectrum between the spectra of neighboring blocks
Therefore, the choice of the optimal spectral distance $\delta$ for gluing the 2 fully optimized QM blocks was carried out proceeding from the simple rule $T(\nu=-\delta)=T(\nu=0)=T(\nu=\delta)$, which reliably guarantees the homogeneity of the entire finite region. 
The choice of the optimal coupling constant $k$ for the case of equidistant frequency comb was carried out on the basis of optimization of the quantity $|T(\nu)-T(0)|^2$ according to \cite{Perminov2017superefficient}, which reduces to solving the equation $\partial_{\nu}^2T(\nu)|_{\nu=0}=0$ with respect to $k$.

As a result, we obtained the detuning $\delta=2,18$ for gluing the optimal QM blocks and $k=4,26$ for the case of 4 resonators. 
As it can be seen in Fig. \ref{fig_3}, the spread of the delay time within the frequency band $[-3\Delta; 3\Delta]$ for the case of 4 resonators is $0,06$, and this value is $0,04$ for two glued universal QM blocks, which corresponds to the best spectral quality of the composite QM. 
A similar procedure for blocks gluing  can be performed for larger number of blocks by using the optimization methods proposed in \cite{Perminov2017spectral}, which makes it possible in addition improving the quality and the working bandwidth of the composite QM.

\textit{Conclusion.}
We proposed a scheme of composite QM allows an effective optimization of the working spectrum for the broadband QM composed from the frequency comb of ring microresonators interacting with each other and with a common waveguide. 
The convenience of controlling the characteristics of the proposed multi-resonator QM is provided by the fact that it consists of pre-optimized spectral characteristics of the three-resonator blocks, each of which can be additionally connected/disconnected independently of the others.
In this case, the procedure for blocks gluing allows us increasing the working bandwidth of  the final composite QM by several orders of magnitude while maintaining high efficiency.

The studied QM scheme can be realized both in the optical and microwave ranges on the basis of existing technologies \cite{Yariv1999,Gorodetsky1999,Armani2003} and is of great interest for the creation of an essentially significant multi-qubit quantum interface and QM for a universal quantum computer. 
A remarkable property of such QM in the optical range is the possibility of its direct use at room temperature due to the high energy of the optical quantum. 
At the same time, an important circumstance is the development of a method for dynamical controlling the storage time of the light field \cite{Sandberg2008,Asfaw2017}. 
This seems possible due to the controllable disconnection of resonators from the waveguide for a given time \cite{Sandberg2008,Pierre2014,Wulschner2016,Flurin2015}, however, the study of this question requires additional research \cite{Brecht_2016,Gu_2017}.

\textit{Acknowledgments.}
This work was financially supported by the Russian
Science Foundation through the Grant No. 14-12-01333-P (basic idea and analysis of results -- SAM, NSP) and was supported in the framework of the budgetary theme of the Laboratory of Quantum Optics and Informatics KPTI SSS FRC KSC RAS (numerical simulation -- NSP, TDY).

\bibliographystyle{apsrev4-1}
\bibliography{CFC_QM}

\end{document}